\documentclass[preprint,superscriptaddress,amssymb,amsmath,aps,longbibliography,prb]{revtex4-2}
\usepackage{graphicx}
\pdfoutput=1

\graphicspath{{figures/}}
\usepackage{color}
\usepackage[colorlinks,bookmarks=false,citecolor=darkblue,linkcolor=red,urlcolor=blue]{hyperref}

\definecolor{darkred}{rgb}{0.7,0.0,0.0}
\definecolor{darkblue}{rgb}{0,0.02,0.45}

\definecolor{darkgreen}{rgb}{0.02,0.45,0.0}

\newcommand{\qv}{\mathbf q}

\begin{document}


\title{Beyond kagome: $p$-bands in kagome metals}

\author{Alexander A. Tsirlin}
\email{altsirlin@gmail.com}
\affiliation{Felix Bloch Institute for Solid-State Physics, University of Leipzig, 04103 Leipzig, Germany}

\author{Ece Uykur}
\email{e.uykur@hzdr.de}
\affiliation{Helmholtz-Zentrum Dresden-Rossendorf, Inst Ion Beam Phys \& Mat Res, D-01328 Dresden, Germany}


\begin{abstract}
We review recent studies on quantum materials where transition-metal atoms give rise to $d$-bands typical of kagome metals. Using examples from several material families -- AV$_3$Sb$_5$, FeGe, RV$_6$Sn$_6$, and LaRu$_3$Si$_2$ -- we argue that $p$-bands contributed by elements beyond the kagome network also play a crucial role in the electronic instabilities, including the charge-density-waves and superconductivity in kagome metals. 
\end{abstract}

\maketitle
\newpage


\section{Introduction}

\begin{figure}[!b]
\includegraphics[width=11cm]{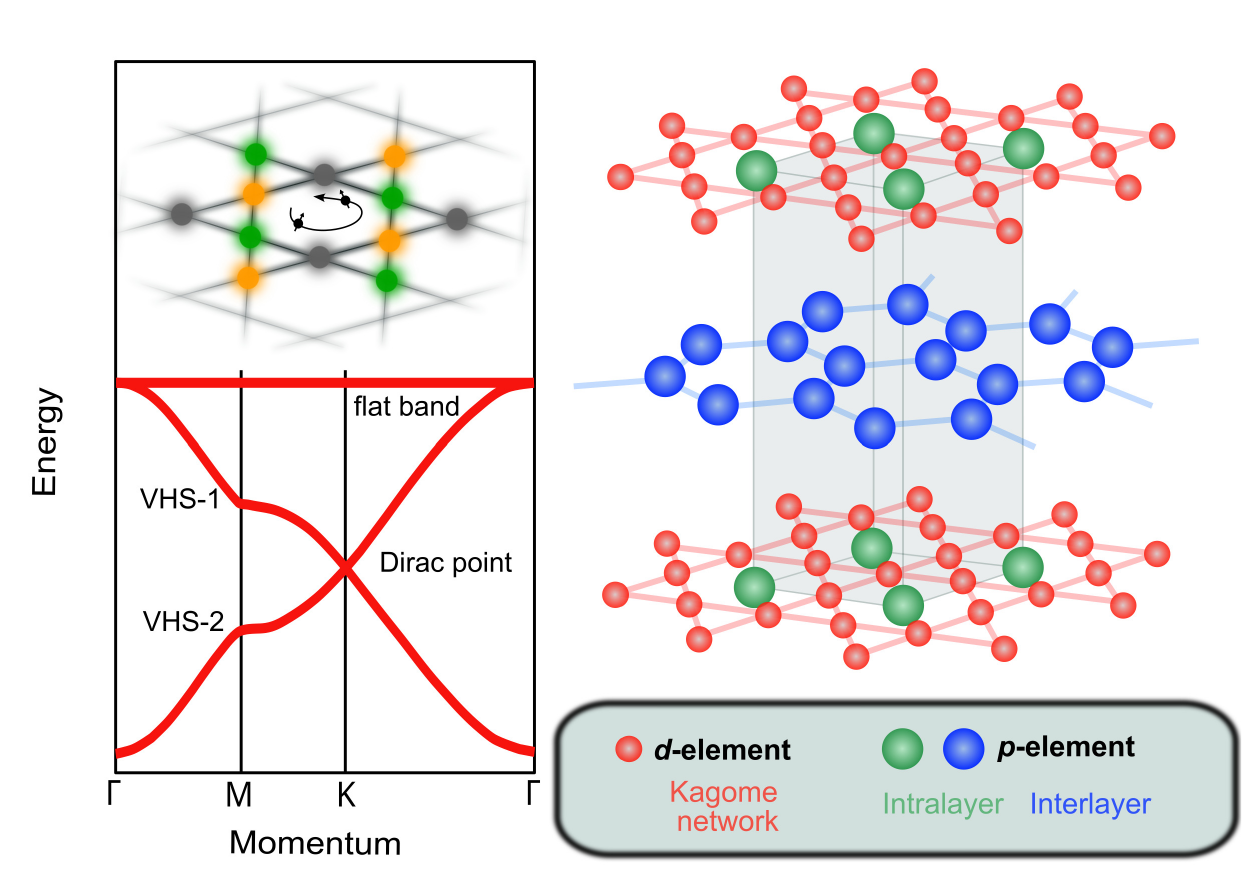}
\caption{\label{fig:intro}
Left: model band structure of a nearest-neighbor kagome metal features a flat band, a Dirac crossing, and two band saddle points that give rise to van Hove singularities (VHS) in the density of states. Right: real crystal structure that combines the kagome network of a $d$-metal with $p$-elements located in the centers of the hexagonal voids (``intralayer'') as well as between the kagome layers (``interlayer'').
}
\end{figure}

Materials with layered triangular networks, the so-called kagome metals, became one of the major topics in solid-state research~\cite{yin2022,wang2023,wang2024}. Their interesting properties are traced back to the effect of destructive interference of nearest-neighbor hoppings on the kagome lattice. This effect gives rise to the flat band, the linear Dirac crossing, and the band saddle points that produce van Hove singularities (VHS) in the density of states (Fig.~\ref{fig:intro}). Adjusting the Fermi level to one of these special points creates various instabilities and leads to unusual electronic properties, thus rendering kagome metals prospective quantum materials.

Kagome networks are typically built by transition-metal atoms that give rise to $d$-bands in the electronic structure. Five $d$-orbitals per atom result in a plethora of bands that nevertheless still show main features of the nearest-neighbor kagome band structure. Interestingly, these numerous $d$-bands form only a part of the electronic system of a real material. Hexagonal voids of the kagome network are rarely empty. They are filled by $p$-elements stabilizing the network. Additionally, $p$-elements necessarily appear between the kagome layers, usually in the form of hexagonal nets (Fig.~\ref{fig:intro}). The effect of these $p$-elements on the band structure is different from the behavior known from ionic compounds, such as transition-metal oxides where the $p$-states of oxygen either lie well below the Fermi level (in Mott insulators) or admix to the $d$-states via metal-oxygen hybridization (in charge-transfer insulators). Kagome metals show a high degree of covalency, in particular, the direct bonding between the $p$-elements. This bonding gives rise to the formation of distinct $p$-bands that will often appear near the Fermi level, produce additional Fermi surfaces, and potentially drive various instabilities concurrent with the native electronic instabilities of the kagome $d$-bands.

In this Perspective, we review several of the recently studied kagome metals and show that many of their properties should be traced back to the $p$-bands present in the material. We focus on systems with structural phase transitions, which are typically interpreted as charge-density wave (CDW) instabilities of the underlying kagome network. We argue that a complete microscopic understanding of these instabilities requires the treatment of $p$-states and $d$-states on equal footing. We also show how the $p$-bands can be instrumental in tailoring unusual electronic properties of the kagome metals.


\section{AV$_3$Sb$_5$ (A = K, Rb, Cs)}

\begin{figure}
\includegraphics[scale=1.2]{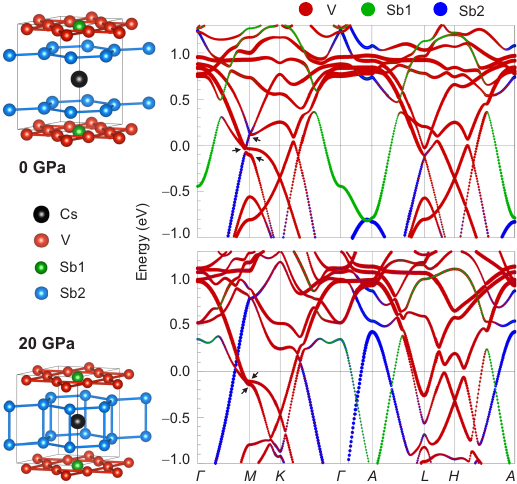}
\caption{\label{fig:pressure}
Pressure-induced 2D--3D crossover in CsV$_3$Sb$_5$~\cite{tsirlin2022}: the shrinkage of the $c$ parameter leads to the formation of the interlayer Sb2--Sb2 bonds and a reconstruction of the band structure that mainly affects the Sb1 and Sb2 bands, whereas the kagome $d$-bands of vanadium show minor changes only. Small arrows denote the band saddle points that give rise to the van Hove singularities near the Fermi level. 
}
\end{figure}

The AV$_3$Sb$_5$ compounds are arguably the most popular family of the kagome metals~\cite{neupert2022}. They have been extensively covered in a recent review article~\cite{wilson2024}, so we focus on the main aspects only. Two competing instabilities, superconductivity below $T_c=1-3$\,K and CDW below $T_{\rm CDW}$ of about $80-100$\,K, are consistently observed in these materials. 
Electronic structure features prominent V $3d$ bands with several VHS at the $M$-point of the Brillouin zone within 100\,meV from the Fermi level (Fig.~\ref{fig:pressure}). 
The respective microscopic models that include the $d$-bands with one or several VHS, augmented by the on-site as well as intersite Coulomb repulsion terms, have been the main stage for the theoretical description of AV$_3$Sb$_5$. However, an orbital decomposition of the band structure also reveals several bands of the Sb $5p$ origin (Fig.~\ref{fig:pressure}). One of these $p$-bands (Sb1) forms a distinct Fermi surface around $\Gamma$. The Sb2 bands hybridize with the V $3d$ bands near $M$ where band saddle points responsible for the VHS are located.

\begin{figure}
\includegraphics[width=13cm]{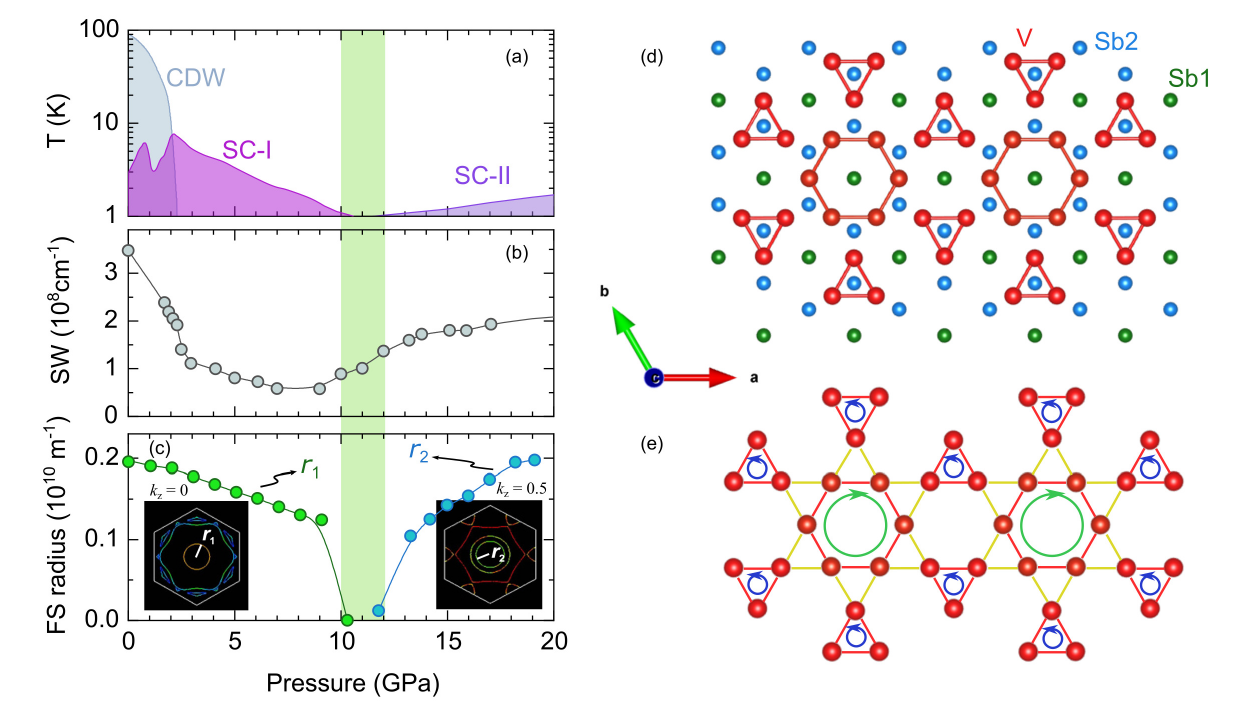}
\caption{\label{fig:av3sb5}
(a-c) Pressure evolution of CsV$_3$Sb$_5$: $T_c$ and $T_{\rm CDW}$ vs. pressure~\cite{zhou2024} (a), spectral weight of the localization peak as an experimental gauge of electron-phonon coupling~\cite{wenzel2023} (b), and radii of the Sb $p$-band Fermi surfaces~\cite{tsirlin2023} (c). Possible CDW structures: real CDW with the modulation of the V--V bond distances (d), and imaginary CDW with the loop currents (e). 
}
\end{figure}

Whereas $T_{\rm CDW}$ of AV$_3$Sb$_5$ is rapidly suppressed by pressure and by (most of the) chemical substitutions, superconductivity remains~\cite{wilson2024}. It shows a highly unusual behavior, for example as a function of pressure where $T_c$ first increases upon the suppression of the CDW, then decreases once CDW has been eliminated, and then increases again at even higher pressures~\cite{zhou2024} (Fig.~\ref{fig:av3sb5}a). This reentrant behavior is generic for all three compounds of the AV$_3$Sb$_5$ family, with critical pressures increasing upon replacing the smaller A = K atom with the larger Rb or Cs.

Pressure evolution of $T_c$ can be traced back to the changes in the electronic structure. Whereas the $d$-bands almost do not change under pressure, with the VHS features below the Fermi level being remarkably intact, the $p$-bands undergo a major reconstruction. The $p$-band Fermi surface near $\Gamma$ completely disappears, and a new Fermi surface around $A$ appears at higher pressures~\cite{tsirlin2022} (Fig.~\ref{fig:pressure}). These changes perfectly correlate with the reentrant behavior of superconductivity, as shown not only by band-structure calculations, but also experimentally (Fig.~\ref{fig:av3sb5}b,c). High-pressure infrared spectroscopy demonstrates the nonmonotonic changes in the intraband response that coincide with the pressure evolution of $T_c$~\cite{wenzel2023}. The origin of this behavior lies in the gradual crossover from the layered 2D structure at ambient pressure to the 3D structure at high pressures. The $c$ parameter shrinks by about 20\% between 0 and 20\,GPa, thus reducing the distance between the hexagonal Sb nets~\cite{tsirlin2022,yu2022}. The Sb--Sb bonds forming between these nets (Fig.~\ref{fig:pressure}) modify the dispersion of the Sb bands and lead to the Fermi surface reconstruction that, in turn, affects the electron-phonon coupling~\cite{wang2022} and, eventually, superconductivity~\cite{ritz2023}.

Back to the ambient-pressure regime, superconductivity emerges from the CDW state, which is highly unconventional. Experimentally, the exotic nature of the CDW in AV$_3$Sb$_5$ is witnessed by chiral responses and by the anomalous Hall effect that appears below $T_{\rm CDW}$ despite the absence of magnetic order, although time-reversal symmetry breaking may occur~\cite{wilson2024}. Microscopically, these observations have been understood as a coexistence of a conventional CDW (charge order of the V--V bonds) and an imaginary CDW represented by an orbital magnetic state with loop currents (Fig.~\ref{fig:av3sb5}d,e). Such a complex pair-density wave instability is indeed anticipated in a kagome metal with the filling level near the VHS. Therefore, typical interpretations of the CDW formation in AV$_3$Sb$_5$ rely on the nesting between the VHS proximate to the Fermi level~\cite{neupert2022}. For example, in CsV$_3$Sb$_5$ these VHS are found both slightly below and slightly above the Fermi level at the $M$-point of the Brillouin zone (Fig.~\ref{fig:pressure}). 

Several observations challenge this scenario. First, the propagation vectors of the CDW [$\qv_1=(\frac12,\frac12,\frac12)$ in KV$_3$Sb$_5$ and RbV$_3$Sb$_5$, and $\qv_2=(\frac12,\frac12,\frac14)$ or a mixture of $\qv_1$ and $\qv_2$ in CsV$_3$Sb$_5$] are not chosen by the Fermi surface nesting~\cite{kaboudvand2022}. Second, the pressure-induced CDW of CsV$_3$Sb$_5$ with the propagation vector $\qv_3=(0,\frac38,\frac12)$ can not be associated with any nesting at all~\cite{stier2024}. Third, strain effect on CsV$_3$Sb$_5$ leads to a shift of the VHS toward the Fermi level, but the CDW is suppressed~\cite{lin2024}.

These observations suggest that the VHS alone do not account for the stability of the CDW in AV$_3$Sb$_5$. Indeed, Sb atoms provide a major contribution to the stabilization energy of the CDW~\cite{ritz2023b}, because any changes in the V--V distances must be accompanied by the displacements of Sb atoms, which are strongly bound to vanadium. Resonant x-ray scattering at the Sb $L_1$ edge witnesses the involvement of Sb $p$-states in the CDW formation~\cite{li2022}. Moreover, the suppression of CDW under pressure can be rationalized if one considers that the shrinkage of the $c$ parameter hinders the out-of-plane displacements of Sb, which are needed to stabilize the CDW state. By contrast, the VHS and their nesting show no appreciable changes under pressure~\cite{tsirlin2022} and can't explain the suppression of the CDW. Therefore, even if CDW is driven by electronic effects in the $d$-bands, it is the $p$-states of Sb that decide on the formation of the CDW, including the choice of its propagation vector and the stabilization energy with respect to the normal state.

One further limitation of the purely $d$-band scenario becomes clear from the comparison between different members of the AV$_3$Sb$_5$ family. While the in-plane lattice parameters of these compounds are almost the same, the $c$-parameter is strongly influenced by the size of the alkaline metal. The $c/a$ ratio decreases from 1.694 in CsV$_3$Sb$_5$ to 1.658 in RbV$_3$Sb$_5$ and 1.634 in KV$_3$Sb$_5$ at ambient pressure~\cite{ortiz2019}. At first glance, one could consider the Rb and K compounds as the compressed versions of CsV$_3$Sb$_5$ corresponding to the nominal pressures of 0.8 and 1.2 GPa, respectively. In this case, the $T_{\rm CDW}$ should be reduced to $40-50$\,K, but in fact it remains at 78\,K in KV$_3$Sb$_5$ and even increases to 104\,K in RbV$_3$Sb$_5$ compared to 94\,K in CsV$_3$Sb$_5$~\cite{wilson2024}. The change in the CDW propagation vector is observed as well and may be related to a different sequence of the VHS in the K and Rb vs. Cs compounds, as confirmed spectroscopically~\cite{wenzel2022}. Further work would be needed to elucidate this change also on the microscopic level, but it clearly requires the consideration of the $p$-states, which are mainly influenced by the change in $c/a$, whereas the kagome network itself remains essentially intact.


\section{FeGe}

\begin{figure}
\includegraphics[scale=1.2]{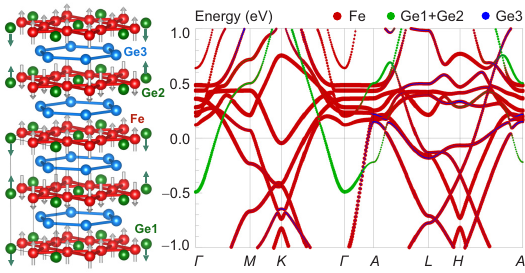}
\caption{\label{fig:fege}
Left: crystal structure of FeGe with the A-type antiferromagnetic order. The green arrows on the Ge atoms show the displacements below $T_{\rm CDW}$. Right: band structure of FeGe in the normal state (A-type AFM order) features Fe $d$-bands along with the $p$-band formed by the Ge1 and Ge2 atoms. 
}
\end{figure}

In contrast to the AV$_3$Sb$_5$ kagome metals discovered only in 2019~\cite{ortiz2019}, hexagonal polymorph of FeGe had been extensively studied already in 1970's and received renewed attention in the recent years in context of the kagome research. Its main difference from all other materials covered in this Perspective is the presence of the A-type antiferromagnetic order (ferromagnetic layers stacked antiferromagnetically, Fig.~\ref{fig:fege}) below 410\,K with a further transition to an incommensurate and canted antiferromagnetic phase below 60\,K~\cite{bernhard1984}. 

A new addition to this already complex behavior is the discovery of yet another transition at 110\,K~\cite{teng2022} interpreted as a formation of a CDW, based on the observation of the $2\times 2$ in-plane superstructure~\cite{teng2022,yin2022b}, the partial gap opening below $T_{\rm CDW}$~\cite{teng2023}, and the presence of VHS near the Fermi level~\cite{teng2023}, all reminiscent of AV$_3$Sb$_5$. However, an optical probe revealed an enhancement of the low-energy spectral weight in FeGe below $T_{\rm CDW}$~\cite{wenzel2024}, at odds with the typical CDW behavior where spectral weight is transferred from lower to higher energies as a result of the gap opening. 

These seemingly conflicting observations could be reconciled using the experimental crystal structure below $T_{\rm CDW}$ where almost no deformation of the Fe kagome layers has been found. The main structural change is due to out-of-plane displacements of the Ge atoms that center the kagome layers~\cite{chen2024a} (Fig.~\ref{fig:fege}). One quarter of these Ge atoms undergoes dimerization that splits the respective Ge $p$-band and accounts for the changes in the electronic structure below $T_{\rm CDW}$~\cite{zhao2025,tan2025}. Concurrently, the $2\times 2$ structural modulation folds the Fe $d$-bands without any significant reconstruction of the Fermi surface~\cite{tang2024} and facilitates low-energy optical transitions that enhance the respective spectral weight~\cite{wenzel2024}. The distortion in the Ge sublattice, preceded by a short-range precursor state above $T_{\rm CDW}$~\cite{subires2025}, imposes a structural modulation on the Fe sublattice without gapping out the respective $d$-bands. This scenario is entirely different from the usual CDW. Indeed, no Kohn anomaly (phonon softening due to electron-phonon coupling) was observed in FeGe. The phonons harden upon cooling through $T_{\rm CDW}$ instead~\cite{miao2023,teng2024}.

The proclivity of the Ge atoms for dimerization appears to be highly sensitive to the distribution of Ge vacancies influenced by post-annealing~\cite{klemm2025}. This subtlety explains several observations of the dimerized state being either long-range-ordered or only short-range-ordered, with dissimilar manifestations of the transition anomalies at $T_{\rm CDW}$ depending on the thermal treatment of the sample~\cite{wu2024,shi2024}. 

The transition at $T_{\rm CDW}$ is mainly driven by a spin-phonon coupling~\cite{miao2023}, the gain in the magnetic energy upon the Ge dimerization~\cite{wang2023b,shao2023}. In contrast to the uncorrelated CsV$_3$Sb$_5$, FeGe is moderately correlated, as seen from the significant renormalization of the band energies in photoemission~\cite{teng2023} and optical~\cite{wenzel2024} spectra compared to density-functional calculations, and further confirmed by a calculation of the on-site Coulomb interactions~\cite{ma2024,jiang2025}. Adding correlations is indeed necessary to soften the relevant phonon mode and facilitate the dimerization transition~\cite{ptok2024,ma2024}, but no such softening happens on approaching $T_{\rm CDW}$ experimentally~\cite{miao2023,teng2024}. The effect of correlations appears to be quantitative and not qualitative, because the low-temperature structure with the dimerized Ge atoms is stable even in the absence of correlations~\cite{miao2023,zhang2024,tan2025}. On the other hand, electronic correlations may still be relevant to some other exotic properties observed in FeGe. Those include the anomalous Hall effect reported below $T_{\rm CDW}$~\cite{teng2022}, the prominent edge states detected by scanning tunneling spectroscopy~\cite{yin2022b}, the unusual increase in the crystalline symmetry on cooling across $T_{\rm CDW}$~\cite{wu2024b}, and the incommensurate spin excitations that appear well above $T_{\rm CDW}$ before condensing into the canted magnetic state below 60\,K~\cite{chen2024b}. The complete microscopic origin of these effects requires further dedicated investigation. 


\section{RV$_6$Sn$_6$ (R = Sc, Y, Lu) and LuNb$_6$Sn$_6$}

\begin{figure}
\includegraphics[scale=1.2]{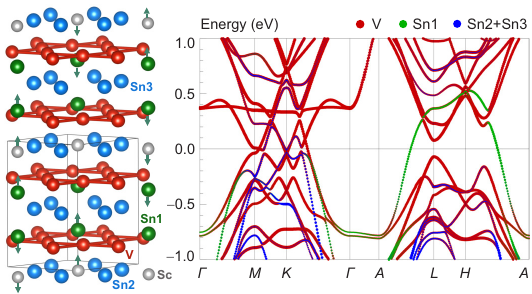}
\caption{\label{fig:scv6sn6}
Left: crystal structure of ScV$_6$Sn$_6$ with the arrows showing the atomic displacements in the CDW state~\cite{arachchige2022}. Right: band structure of ScV$_6$Sn$_6$ (normal state) near the Fermi level features multiple V $3d$ bands and a separate band with the dominant Sn1 contribution.
}
\end{figure}

Nonmagnetic ScV$_6$Sn$_6$ offers an interesting comparison to AV$_3$Sb$_5$ because it shows a CDW with the similar $T_{\rm CDW}$ of 92\,K, yet no superconductivity~\cite{arachchige2022}. Electronic structure of this compound features multiple VHS of the V $d$-bands in the vicinity of the Fermi level~\cite{hu2024,cheng2024} (Fig.~\ref{fig:scv6sn6}), thus reinforcing the similarity to AV$_3$Sb$_5$ and inspiring a similar microscopic interpretation. However, the CDW of ScV$_6$Sn$_6$ adopts a different propagation vector of $\qv=(\frac13,\frac13,\frac13)$~\cite{arachchige2022} that has not been observed in the AV$_3$Sb$_5$ family. The CDW state is mainly characterized by the out-of-plane displacements of the Sc and Sn atoms~\cite{arachchige2022,korshunov2023,hu2025} (Fig.~\ref{fig:scv6sn6}). 

The bilayer structure of ScV$_6$Sn$_6$ gives rise to the Sn $p$-band crossing the Fermi level (Fig.~\ref{fig:scv6sn6}), and mostly these Sn states are affected by the CDW formation~\cite{lee2024}, whereas the $d$-band VHS are unchanged below $T_{\rm CDW}$~\cite{kim2023}, and no gap opening is seen in the optical spectra~\cite{hu2023}. Electron-phonon coupling associated with the Sn $p$-orbitals was identified as the primary mechanism of this instability~\cite{korshunov2023,hu2025} that can also be seen as an order-disorder transition, because significant CDW correlations are observed above $T_{\rm CDW}$~\cite{pokharel2023,alvarado2024} and even cause a peculiar electronic response reminiscent of a pseudogap~\cite{destefano2023}. Above $T_{\rm CDW}$, ScV$_6$Sn$_6$ is characterized by a close competition between the $\qv=(\frac13,\frac13,\frac13)$ and $\qv=(\frac13,\frac13,\frac12)$ instabilities arising from the flat phonon band in this material~\cite{korshunov2023,hu2025,pokharel2023,alvarado2024,cao2023,tan2023}.

Interestingly, already 5\% of the Y or Lu substitution on the Sc site leads to a complete suppression of the CDW~\cite{meier2023}, whereas pure YV$_6$Sn$_6$ and LuV$_6$Sn$_6$ do not show any CDW either~\cite{pokharel2021,mozaffari2024}. These observations were explained by the geometrical effect of the small Sc atoms that allow the out-of-plane displacements of Sn1, whereas larger Y and Lu atoms prevent the Sn displacements and hinder the CDW formation. Consistent with these trends, external pressure eliminates the CDW above 2.4\,GPa~\cite{zhang2022}, whereas the in-plane compressive strain with the concomitant elongation of the $c$ parameter stabilizes the CDW compared to the strain-free sample~\cite{tuniz2025}. From crystal structure standpoint, the CDW in ScV$_6$Sn$_6$ shows intriguing similarities to FeGe where the CDW instability is also caused by the $p$-element located in the center of the kagome network with almost no impact on the kagome network itself. 

In-depth studies of the CDW state of ScV$_6$Sn$_6$ revealed further peculiarities, including the anomalous Hall effect~\cite{mozaffari2024,yi2024}, possible time-reversal symmetry breaking~\cite{guguchia2023}, and various nematic instabilities~\cite{jiang2024,farhang2025}. The current theoretical interpretation of these phenomena hinges upon the presence of multiple VHS near the Fermi level and the ensuing excitonic order~\cite{ingham2024}. The underlying assumption is that the Sn $p$-states become gapped below $T_{\rm CDW}$ and expose the physics of the V $d$-bands. On the other hand, ScV$_6$Sn$_6$ also features competing CDW instabilities~\cite{korshunov2023,pokharel2023,alvarado2024,cao2023,tan2023} and strongly anharmonic phonons~\cite{wang2024b} that are likely to affect the behavior of the material also below $T_{\rm CDW}$. The role of these effects remains to be explored in future theoretical studies.

Recent work on the isostructural Nb-based kagome metals RNb$_6$Sn$_6$ (R = rare-earth) uncovered striking similarities to the RV$_6$Sn$_6$ family discussed above. Among the Nb family, only LuNb$_6$Sn$_6$, the compound with the smallest R-cation, shows a CDW instability~\cite{ortiz2025}. It features the same CDW propagation vector as in ScV$_6$Sn$_6$~\cite{ortiz2025}, and mostly the Sn $p$-states are affected by the CDW formation, whereas the Nb $d$-states remain largely intact~\cite{lou2025}. External pressure reduces the Lu--Sn distances and suppresses the CDW already above 1.9\,GPa~\cite{meier2025}. All these observations reinforce the scenario of the Sn atoms and their $p$-states causing the CDW instability in the 166 kagome metals.


\section{LaRu$_3$Si$_2$}

\begin{figure}
\includegraphics[scale=1.2]{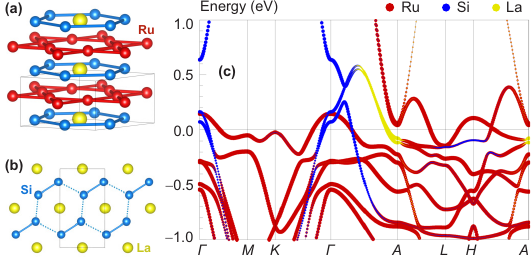}
\caption{\label{fig:laru3si2}
(a) Hexagonal crystal structure of LaRu$_3$Si$_2$ above 600\,K ($P6/mmm$). (b) Orthorhombic crystal structure below 600\,K ($Cccm$) shows the primary deformation mode due to in-plane displacements of Si, resulting in the dimerization of the honeycomb network~\cite{plokhikh2024}. (c) Band structure of LaRu$_3$Si$_2$ calculated using the high-temperature ($P6/mmm$) crystal structure for an easier comparison with other materials. 
}
\end{figure}

Similar to AV$_3$Sb$_5$, LaRu$_3$Si$_2$ manifests both CDW and superconductivity. It belongs to a large family of superconducting 132 compounds with the highest $T_c$ of 7\,K at ambient pressure. Doping dependence of the $T_c$~\cite{li2016,chakrabortty2023} along with muon spectroscopy experiments~\cite{mielke2021} consistently indicate nodeless superconductivity in the parent compound, whereas Fe doping renders superconductivity nodal before completely suppressing it at about 4\% substitution~\cite{mielke2024}. Calculations based on the electron-phonon coupling significantly underestimate the $T_c$~\cite{mielke2021}, indicating a potentially unconventional mechanism of the superconductivity. Indeed, heat capacity of LaRu$_3$Si$_2$ deviates from the typical metallic behavior suggesting the effect of electronic correlations~\cite{li2011,mielke2024}. Such correlations are usually associated with the presence of weakly dispersive $d$-bands near the Fermi level. Additionally, VHS typical of a kagome metal appear in the band structure near the $M$-point (Fig.~\ref{fig:laru3si2}). 

The superconductivity in LaRu$_3$Si$_2$ is intertwined with two concurrent CDW-like instabilities. Below $T_{\rm CO,1}\simeq 400$\,K, a structural modulation with $\qv_1=(\frac14,0,0)$ appears, followed by another structural modulation with $\qv_2=(\frac16,0,0)$ below $T_{\rm CO,2}\simeq 80$\,K, and an additional transition at $T^*\simeq 35$\,K of hitherto unknown origin~\cite{mielke2024b}. The increase in the magnetoresistance below $T_{\rm CO,2}$~\cite{mielke2024b} evokes parallels to the anomalous Hall effect observed in the CDW states of the nominally nonmagnetic AV$_3$Sb$_5$ and ScV$_6$Sn$_6$. However, in contrast to AV$_3$Sb$_5$, the superconducting and putative CDW instabilities in LaRu$_3$Si$_2$ are synergistic because they show similar evolution under pressure. The superconducting $T_c$ remains around 7\,K up to 12\,GPa and decreases at higher pressure concomitant with the transformation of the long-range-ordered superstructure into short-range charge correlations~\cite{ma2024b}. Negative pressure introduced by Ge doping on the Si site increases $T_c$~\cite{misawa2025}. 

Spectroscopic experiments would be required to explore the changes in the electronic structure below $T_{\rm CO,1}$ and $T_{\rm CO,2}$ and the nature of the electronic states that become gapped in the putative CDW phases. Importantly, though, the normal state of LaRu$_3$Si$_2$ features strongly intertwined $d$- and $p$-bands, similar to the other kagome metals. A dispersive Si $3p$ band is clearly seen in the band structure along with a sizable contribution from La (see Fig.~\ref{fig:laru3si2}), which centers the honeycomb network of Si atoms and participates in the bonding.

Owing to the larger size of Ru, the in-plane lattice parameter of LaRu$_3$Si$_2$ is increased compared to $3d$ kagome metals. This in-plane stretching results in longer Si--Si distances in the honeycomb network and renders it unstable. Indeed, already at 600\,K, above $T_{\rm CO,1}$, LaRu$_3$Si$_2$ undergoes another structural phase transition with the in-plane dimerization of the Si atoms (Fig.~\ref{fig:laru3si2}) and only a minor impact on the kagome layers~\cite{plokhikh2024}. The transition at $T_{\rm CO,1}$ renders the kagome layer more deformed and slightly buckled, but it also involves additional displacements of Si~\cite{plokhikh2024}, thus making the $p$-states and associated instabilities a potentially important ingredient of the LaRu$_3$Si$_2$ physics as well. Note that LaRu$_3$Si$_2$ does not feature Si atoms in the kagome layers. Therefore, the primary displacement mode of the $p$-element is in-plane, in contrast to the out-of-plane modes in ScV$_6$Sn$_6$ and FeGe. 


\section{Concluding remarks}

\begin{figure}
\includegraphics[width=11cm]{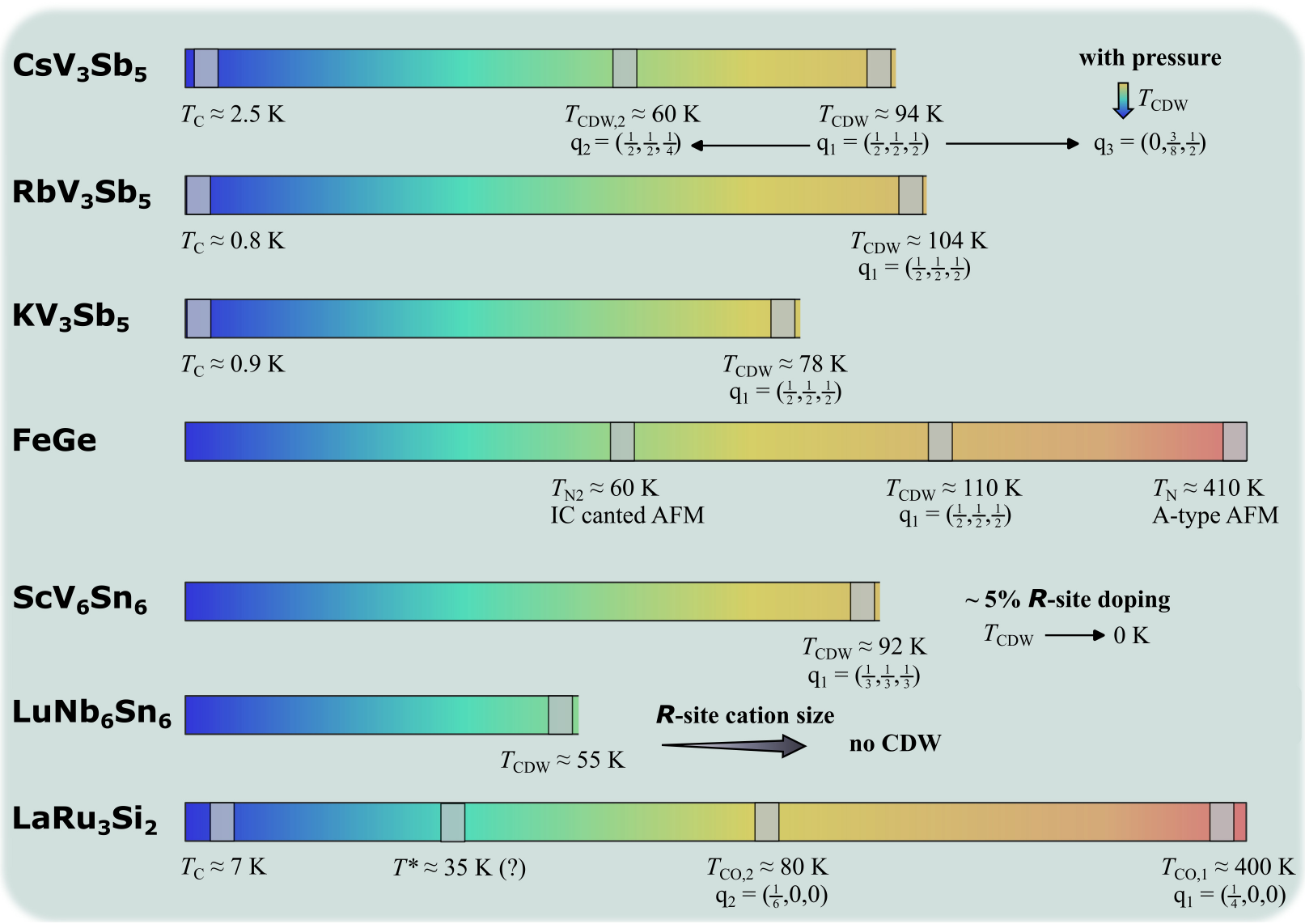}
\caption{\label{fig:summary}
Summary of the kagome metals reviewed in this Perspective. The critical temperatures of the superconducting and CDW instabilities are indicated, along with the CDW propagation vectors. 
}
\end{figure}

The kagome metals covered in this Perspective are quite different in terms of their energy scales and ground states (Fig.~\ref{fig:summary}), yet they share several commonalities. All these materials feature van Hove singularities that arise from the $d$-bands and reflect the underlying kagome structure. They also entail distinct $p$-bands that cross the Fermi level and contribute parts of the Fermi surface. In AV$_3$Sb$_5$, the CDW instability gaps out the $d$-bands near the $M$-point of the Brillouin zone, while the $p$-bands remain largely intact. An opposite situation is realized in FeGe and ScV$_6$Sn$_6$ where mainly the $p$-bands become gapped below $T_{\rm CDW}$, whereas the $d$-bands do not. Regardless of the exact scenario, the CDW's of kagome metals are quite unusual because a significant part of the Fermi surface is not affected by the CDW transition. The resistivities of AV$_3$Sb$_5$~\cite{wilson2024}, FeGe~\cite{shi2024}, and ScV$_6$Sn$_6$~\cite{arachchige2022} all decrease below $T_{\rm CDW}$, likely because of the reduced interband scattering. This behavior is at odds with the conventional CDW scenario (increased resistivity due to gap opening) and serves as a fingerprint of the underlying multi-band scenario with a combination of the $d$- and $p$-bands. No structural deformation gaps both of these bands simultaneously. 

The variations of $T_{\rm CDW}$ with hydrostatic, uniaxial, and chemical pressure can not be accounted for by the changes in the $d$-band van Hove singularities. Instead, the evolution of CDW in AV$_3$Sb$_5$, FeGe, and ScV$_6$Sn$_6$ is well explained when the $p$-elements are taken into account. The out-of-plane displacements of these $p$-elements are naturally controlled by the interlayer spacing and allow a strong coupling of the CDW to the external stimuli. The $d$-bands may create electronic pre-conditions for the CDW formation, but the eventual structural distortion and its energy/temperature scale are decided by the $p$-element. Generically, the 2D kagome network should not be strongly affected by pressure, because in-plane hoppings are rather insensitive to the out-of-plane lattice parameter and $c/a$ ratio. The strong pressure response is rooted in the $p$-elements that change their position (AV$_3$Sb$_5$) or proclivity for dimerization (FeGe, ScV$_6$Sn$_6$) with the reduction in $c/a$. LaRu$_3$Si$_2$ can be an interesting reference case here, because its leading atomic displacements are in-plane, also for the $p$-elements. Detailed studies of the CDW's in this compound would be very insightful. 

A natural question at this juncture is whether kagome metals without the $p$-bands are possible at all. In fact, they do exist. Fe$_3$Sn$_2$, Ni$_3$In, RMn$_6$Sn$_6$ (R = rare-earth) all lack distinct $p$-bands in the vicinity of the Fermi level, although they do show some admixture of the $p$-states to the predominant $d$-bands. Incidentally, none of these kagome metals exhibits CDW transitions or any other structural instabilities. This further supports the idea that the $p$-states must be involved -- namely, form distinct bands and parts of the Fermi surface -- for a CDW in a kagome metal to occur. An interesting case for future work is the ATi$_3$Bi$_5$ family (A = Rb, Cs) where Bi $6p$ bands make a prominent contribution near the Fermi level~\cite{wenzel2025}, no CDW is observed, but signatures of electronic nematicity have been reported~\cite{hu2024b}. 

One of the most intriguing features of the kagome metals is their unconventional superconductivity that, at least theoretically, emerges from the various density-wave orders in the kagome $d$-bands~\cite{kiesel2013}. The ambient-pressure superconductivity of AV$_3$Sb$_5$ indeed shows several highly unusual features, such as the abnormal charge-$4e$ and $6e$ flux quantization~\cite{ge2024} and time-reversal symmetry breaking~\cite{deng2024}. From the materials perspective, it may be important that many of the carriers reside in the $p$-band of AV$_3$Sb$_5$, whereas the $d$-bands are at least partially gapped by the CDW. The disappearance of the $p$-band upon compression or doping is clearly correlated with the suppression of superconductivity~\cite{wilson2024,tsirlin2022}. The situation here may be somewhat similar to the CDW: the $p$-bands are responsible for the stabilization of superconductivity, whereas the $d$-bands impart more exotic features to it. Including the $p$-bands within realistic theoretical models would be highly desirable in order to shed further light on this peculiar interplay. The $p$-states may also be relevant to various subtleties of the AV$_3$Sb$_5$ behavior, such as the differences between the strained and strain-free samples~\cite{guo2024} or between the bulk samples and thin flakes~\cite{song2023}, because external stimuli are more likely to couple to the $p$-states than to the kagome network itself.

From the materials design perspective, the $p$-elements of kagome metals offer an excellent opportunity to control the CDW. Chemical substitutions in the $p$-sublattice and even in the A-sublattice, as in the case of ScV$_6$Sn$_6$, can be used to suppress CDW with only a minor impact on the kagome $d$-bands. An ultimate example of this control is FeGe where the appearance of the structural instability and its long-range vs. short-range nature hinge upon the distribution of Ge vacancies in the sample. A periodic drive affecting the $p$-elements, for example via exciting a phonon mode related to these atoms, may be another interesting way of controlling the CDW in kagome metals.

In summary, the $p$-elements and their corresponding $p$-bands are an integral part of kagome metals. They enrich the physics of these quantum materials and offer ample opportunities for controlling the electronic states therein.

\section*{Technical notes}
Band structures shown in this work are calculated with the \texttt{FPLO} code~\cite{fplo} on the full-relativistic level using the Perdew-Burke-Ernzerhof exchange-correlation potential~\cite{pbe96} and experimental structural parameters. \texttt{VESTA} was used for crystal structure visualization~\cite{vesta}.

\acknowledgments
We thank Maxim Wenzel and Martin Dressel for the long-term collaboration on the optical studies of kagome metals, as well as Brenden Ortiz, Stephen Wilson, Claudia Felser, David Mandrus, and Hechang Lei for providing the crystals of these interesting materials. We are also grateful to the Deutsche Forschungsgemeinschaft (DFG, German Research Foundation) for the funding under TRR\,360 (492547816) and UY63/2-1.


%

\end{document}